\documentclass[preprint,11pt,3p]{elsarticle}
\usepackage{amssymb,amsmath,graphicx,bbm}
\begin{document}
\title{The effect of channel decoherence on entangled coherent
states: a theoretical analysis}
\author{Yao Yao}
\author{Hong-Wei Li}
\author{Zhen-Qiang Yin\corref{cor1}}
\ead{yinzheqi@mail.ustc.edu.cn}
\author{Guang-Can Guo}
\author{Zheng-Fu Han\corref{cor2}}
\ead{zfhan@ustc.edu.cn}
\cortext[cor1]{Corresponding author}
\cortext[cor2]{Principal corresponding author}
\address{Key Laboratory of Quantum Information,University of Science and
Technology of China,Heifei 230026,China}
\date{\today}

\begin{abstract}
We analyse decoherence properties of entangled coherent states (ECSs) due to channel
losses, and compare to the performance of bi-photon Bell states. By employing the concept
of "entanglement of formation" (EOF), the degradation of fidelity and degree of entanglement
are calculated. Two situations are considered: (1) asymmetric and (2) symmetric noise channel.
In the first case, only the lower bound of EOF is given. However, in the latter case, we have obtained
an explicit expression of concurrence (and then EOF) and found surprisingly our result is just incompatible with that of
[Phys. Rev. A \textbf{82}, (2010) 062325] in the limit of $\alpha\rightarrow0$. We demonstrate
that entangled coherent states with sufficient small amplitudes are more robust against channel decoherence
than biphoton entangled states in both cases.
\end{abstract}


\maketitle


\section{Introduction}
Up to date, quantum entanglement has played a very important role in quantum
information processing (QIP) such as quantum teleportation \cite{teleportation}, quantum computation \cite{computation},
dense coding \cite{dense coding} and quantum key distribution \cite{QKD}. Beside the entangled orthogonal states (e.g., biphoton Bell
states), the entangled nonorthogonal states (e.g., ECSs \cite{sanders}) also attract a lot of attention in
various applications in QIP \cite{QIP1,QIP2,QIP3,QIP4,QIP5,QIP6,QIP7,QIP8,QIP9,QIP10}.
\par
First, we make a brief review of the theory of "entanglement of formation" \cite{EOF1,EOF2,EOF3}. For each pure
entangled bipartite state $\left|\psi\right\rangle_{AB}$, the measure of entanglement is defined as the
entropy of either of the two subsystem A and B:
\begin{equation}
\label{entropy}
E\left(|\psi\rangle_{AB}\right)=-Tr_A\left(\rho_Alog_2\rho_A\right)=-Tr_B\left(\rho_Blog_2\rho_B\right),
\end{equation}
\begin{equation}
\rho_A=Tr_B|\psi\rangle_{AB}\langle\psi|, \rho_B=Tr_A|\psi\rangle_{AB}\langle\psi|.
\end{equation}
For mixed states $\rho=\sum\limits_i{p_i\left|\psi_i\rangle\langle\psi_i\right|}$,
the entanglement of formation is defined as the average entanglement of the pure states
of the decomposition, minimized over all decompositions of  $\rho$:
\begin{equation}
E(\rho)=\min\sum\limits_i{p_iE(\psi_i)}.
\end{equation}
\par
One can show that for pure states, the entanglement defined in Eq. (\ref{entropy}), can be rewritten as:
\begin{equation}
E(\psi)=\mathcal{E}(C(\psi)),
\end{equation}
where the "concurrence" C is defined as $C(\psi)=\left|\langle\psi|\tilde{\psi}\rangle\right|$
with $\left|\tilde{\psi}\right\rangle=\sigma_y\left|\psi\right\rangle^{\ast}$, and the function
$\mathcal{E}$ is given by:
\begin{equation}
\mathcal{E}(C)=H\left(\frac{1+\sqrt{1-C^2}}{2}\right),
\end{equation}
\begin{equation}
H(x)=-x\log_2x-(1-x)\log_2(1-x).
\end{equation}
The above formula still holds for mixed states of bipartite systems \cite{EOF2}.
\par
In this paper we analyse decoherence properties of entangled coherent states (ECSs) due to channel
losses. In the discussion, two situations are considered: one is that only one mode
of ECSs suffers from losses, and the other is that both modes are equally lossy. By employing the concept
of "entanglement of formation", the degradation of fidelity and degree of entanglement are calculated. We found
that in the limit of small amplitudes $\alpha$ entangled coherent states show more robust against decoherence
than the biphoton entangled states.
\section{Entangled coherent states and decoherence model}
A coherent state $\left|\alpha\right\rangle$, defined as an eigenstate of the annihilation operator $\hat{a}$,
can be represented in the photon number states:
\begin{equation}
\left|\alpha\right\rangle=e^{-\left|\alpha\right|^2/2}\sum_{n=0}^{\infty}{\frac{\alpha^n}{\sqrt{n!}}|n\rangle},
\end{equation}
where $\alpha$ is a complex amplitude and the overlap is $\left\langle\alpha|-\alpha\right\rangle=exp(-2|\alpha|^2)$.
In this paper, we discuss the entangled coherent states of the form:
\begin{equation}
\label{ECS}
\left|\psi^{\pm}\right\rangle_{1,2}=\frac{1}{\sqrt{N_{\pm}}}(|\alpha\rangle_1|-\alpha\rangle_2\pm
|-\alpha\rangle_1|\alpha\rangle_2).
\end{equation}
where $N_{\pm}$ is the normalization factor, $N_{\pm}=2\pm2exp(-4\left|\alpha\right|^2)$. One
can see \cite{review} for a review of entanglement properties of ECSs.
\par
To investigate the effect of channel decoherence on entangled coherent states, we can model such photon
losses by interacting the signal with a vacuum mode $|0\rangle_E$ in a beam splitter with
the properly chosen transmissivity parameter $\eta$ (which is equal to the channel noise parameter).
In the remainder of the paper, we will discuss decoherence properties of the state
$|\psi^{-}\rangle$ by utilizing the above error model.

\section{Analysis of decoherence properties of ECSs}
Employing the noisy channel model described in Section 2, the effect of the decoherence
can be expressed as follows:
\begin{equation}
\left|\alpha\right\rangle_{1(2)}\left|0\right\rangle_E\rightarrow
\left|\sqrt{\eta}\alpha\right\rangle_{1(2)}|\sqrt{1-\eta}\alpha\rangle_E
\end{equation}
where $\left|0\right\rangle_E$ refers to the environment mode and $\eta$ is the noise parameter
which means the fraction of photons that \textbf{survive} the noise. In the later discussion,
two situations are considered: one is that only one mode of ECSs suffers from losses,
and the other is that both modes are equally lossy.
\\
\textbf{Case 1: The asymmetric noise channel}\\
In this situation, we suppose that mode 1 is kept by Alice at her location remained unchanged, meanwhile
mode 2 is transferred to Bob. Obviously, mode 2 inevitably suffers from photon losses. If we choose the
odd ECSs $\left|\psi^{-}\right\rangle_{1,2}$ as the initial state, which is a maximally entangled state
regardless of $\alpha$, then the whole state develops into:
\begin{align}
\hfill\left|\psi^{-}\right\rangle_{1,2}\otimes|0\rangle_E\rightarrow
\frac{1}{\sqrt{N_{-}}}(|\alpha\rangle_1|-\sqrt{\eta}\alpha\rangle_2|-\sqrt{1-\eta}\alpha\rangle_E\hfill\nonumber\\
\hfill-|-\alpha\rangle_1|\sqrt{\eta}\alpha\rangle_2|\sqrt{1-\eta}\alpha\rangle_E),\hfill
\end{align}
Without losing generality, $\alpha$ is assumed to be real for simplicity throughout the paper.
If we trace out the environment mode E, the reduced density matrix can be obtained:
\begin{align}
\rho_{1,2}&=
\frac{1}{N_{-}}\{|\alpha\rangle_1\langle\alpha|\otimes|-\sqrt{\eta}\alpha\rangle_2\langle-\sqrt{\eta}\alpha|
+|-\alpha\rangle_1\langle-\alpha|\otimes|\sqrt{\eta}\alpha\rangle_2\langle\sqrt{\eta}\alpha|\nonumber\\
&- \Gamma_1(|\alpha\rangle_1\langle-\alpha|\otimes|-\sqrt{\eta}\alpha\rangle_2\langle\sqrt{\eta}\alpha|
+|-\alpha\rangle_1\langle\alpha|\otimes|\sqrt{\eta}\alpha\rangle_2\langle-\sqrt{\eta}\alpha|)\}.
\end{align}
where $\Gamma_1=exp\{-2(1-\eta)\alpha^2\}$.
This decohered state becomes a mixed state, so it is usually difficult to get the exact amount of entanglement
of formation. To estimate the lower bound of $E(\rho)$, a quantity $f(\rho)$ which called the "fully entangled
fraction" is introduced \cite{EOF1}:
\begin{equation}
f(\rho)=\max\left\langle e|\rho|e\right\rangle ,
\end{equation}
where the maximum is over all completely entangled states $|e\rangle$. The lower bound $E(\rho)\geq h[f(\rho)]$
is defined as \cite{EOF1}:
\begin{eqnarray}
h[f(\rho)]=\left\{\begin{array}{cc}
H[\frac{1}{2}+\sqrt{f(1-f)}] & for \, f\geq\frac{1}{2}\\
0                            & for \, f<\frac{1}{2}
\end{array}\right.
\label{bound}
\end{eqnarray}
Here we follow the same procedure to compute $f(\rho_{1,2})$:
\begin{equation}
f(\rho_{1,2})=\max_{\beta}\left\langle\psi^{-}(\beta)|\rho_{1,2}|\psi^{-}(\beta)\right\rangle,
\end{equation}
where $|\psi^{-}(\beta)\rangle_{1,2}$ is of the similar form of $|\psi^{-}(\alpha)\rangle_{1,2}$,
so $f(\rho_{1,2})$ reads:
\begin{align}
f(\rho_{1,2})&=\max_{\beta}\frac{(1+\Gamma_1)}{2(1-e^{-4|\alpha|^2})(1-e^{-4|\beta|^2)}}
(e^{-|\beta-\alpha|^2}e^{-|\beta-\sqrt{\eta}\alpha|^2}\nonumber\\
&+e^{-|\beta+\alpha|^2}e^{-|\beta+\sqrt{\eta}\alpha|^2}
-2e^{-|\beta-\alpha|^2/2}e^{-|\beta+\alpha|^2/2}e^{-|\beta-\sqrt{\eta}\alpha|^2/2}e^{-|\beta+\sqrt{\eta}\alpha|^2/2}),
\end{align}
The problem is to find the maximal value of $f(\rho_{1,2})$. It is easy to verify that
$\frac{d}{d\beta}f(\rho_{1,2})=0$
when $\beta=\frac{\alpha+\sqrt{\eta}\alpha}{2}$, that is ,the maximum overlap can be reached if:
\begin{equation}
\beta=\frac{\alpha+\sqrt{\eta}\alpha}{2} \,.
\end{equation}
\par
\begin{figure}
\begin{center}
\includegraphics[width=0.41\textwidth]{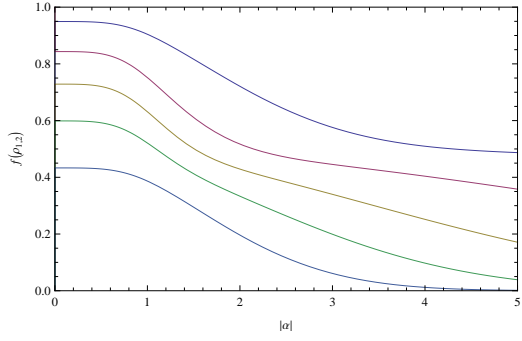}
\end{center}
 \caption{Fully entangled fraction $f(\rho_{1,2})$ as a function of $|\alpha|$
for various noise parameters $\eta=0.9,0.7,0.5,0.3,0.1$  from top to bottom.
   }
\label{f1}
\end{figure}
\begin{figure}
\includegraphics[width=0.41\textwidth]{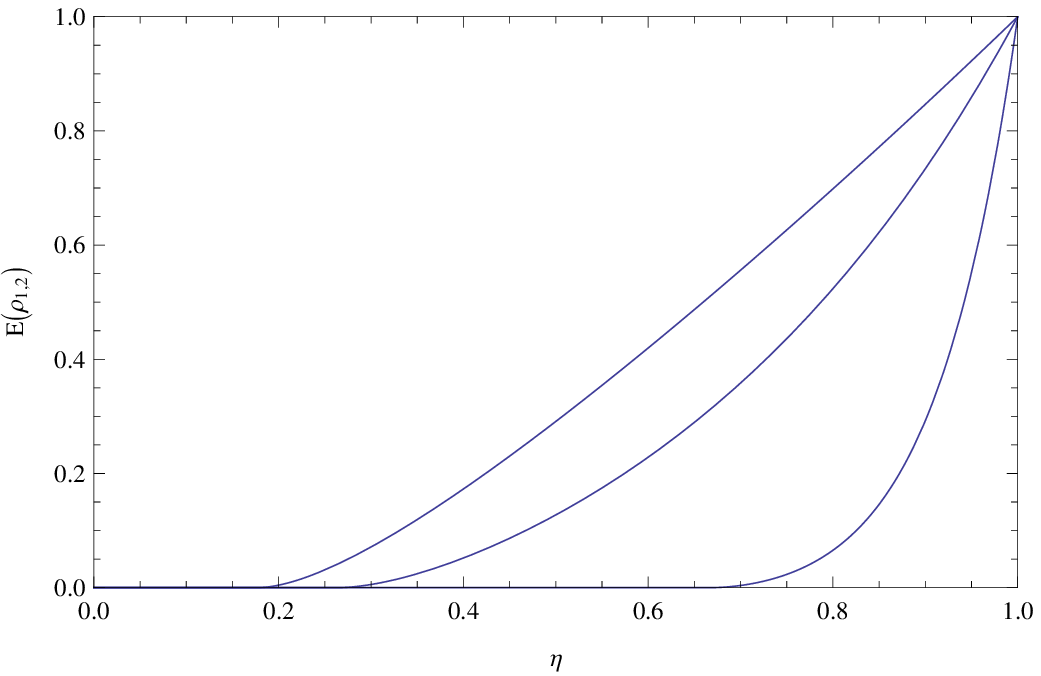}
$\quad$
\includegraphics[width=0.41\textwidth]{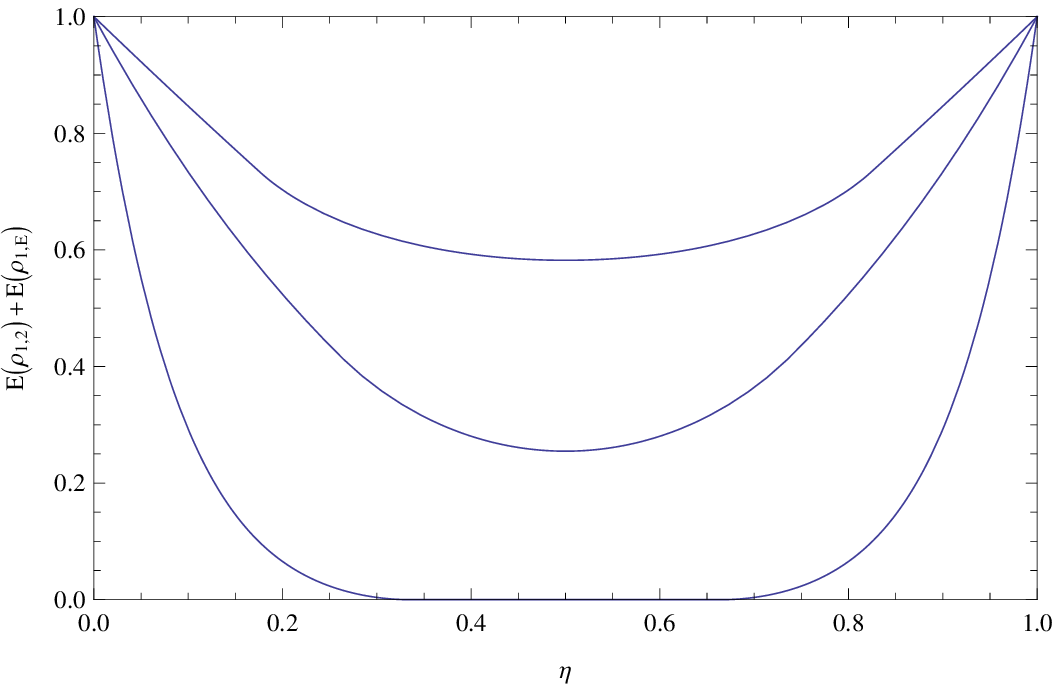}
\caption{The lower bounds of $E(\rho_{1,2})$ and $E=E(\rho_{1,2})+E(\rho_{1,E})$
($E(\rho_{1,E})$ denotes the entanglement between mode 1 and environment mode E)
as a function of the noise parameter $\eta$ for $\alpha=0.5,1.0,2.0$ from top to bottom
(Note that one can directly get $E(\rho_{1,E})$ by substituting $\eta$ for$1-\eta$
in the whole calculation, so the symmetry of E is easy to be understood).}
\label{E1}
\end{figure}
\begin{figure}
\begin{center}
\includegraphics[width=0.41\textwidth]{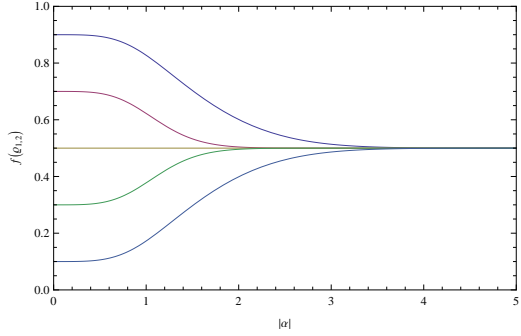}
\end{center}
 \caption{The entanglement fidelity $f(\varrho_{1,2})$ as a function of $|\alpha|$ for
different parameters $\eta=0.9,0.7,0.5,0.3,0.1$ from top to bottom.}
\label{f2}
\end{figure}

Now we can plot the "fully entangled fraction" $f(\rho_{1,2})$ as a function of $|\alpha|$
for various noise parameters $\eta$ in Fig. \ref{f1}.
Comparing with ECSs, we now consider a biphoton Bell state:
\begin{equation}
\label{Bell}
|\psi^{Bell}\rangle_{1,2}=\frac{1}{\sqrt{2}}(|H\rangle_1|V\rangle_2-|V\rangle_1|H\rangle_2),
\end{equation}
where $|H\rangle$ and $|V\rangle$ refer to horizontal and vertical polarization states respectively.
The density matrix of state (\ref{Bell}) after traveling through the same noisy channel becomes:
\begin{equation}
\label{Bell1}
\rho_{1,2}^{Bell}=\eta|\psi^{Bell}\rangle_{1,2}\langle\psi^{Bell}|
+\frac{1}{2}(1-\eta)I_{1}\otimes|0\rangle_{2}\langle0|,
\end{equation}
where $I_{1}$ is the identity matrix of mode 1. So the fully entangled fraction of
$\rho_{1,2}^{Bell}$ is:
\begin{equation}
f(\rho_{1,2}^{Bell})=\eta \,.
\end{equation}
In Fig. \ref{f1}, it is straightforward to see that $f(\rho_{1,2}^{Bell})<f(\rho_{1,2})$ when
$\alpha$ is sufficiently small. Though we can not give the accurate expression of $E(\rho_{1,2})$
in this case, based on the fully entangled fraction $f(\rho_{1,2})$, the lower
bound of entanglement is illustrated in Fig. \ref{E1}.
Intuitively, One can see from Fig. \ref{E1} that the ECSs decohere faster as their
amplitudes become larger, and in the limit of $\alpha\rightarrow\infty$ the entanglement
between mode 1 and 2 will rapidly disappear (except for $\eta=1$ which means lossless).
These facts will be explicitly discussed in the next situation.
\\
\textbf{Case 2: The symmetric noise channel}\\
In this situation, we suppose that mode 1 and 2 are both traveling through the channel
and equally lossy. We will introduce two auxiliary environment modes E1 and E2, which
coupled with mode 1 and 2 respectively, and the initial state reads $|\psi^{-}\rangle_{1,2}\otimes
|0\rangle_{E1}|0\rangle_{E2}$. After passing the noisy channel, we can get the final state as:
\begin{eqnarray}
|\psi^{-}\rangle_{1,2}\otimes
|0\rangle_{E1}|0\rangle_{E2}
&\rightarrow\frac{1}{\sqrt{N_{-}}}(|\sqrt{\eta}\alpha\rangle_1|-\sqrt{\eta}\alpha\rangle_2|\sqrt{1-\eta}\alpha\rangle_{E1}
|-\sqrt{1-\eta}\alpha\rangle_{E2}\nonumber \\
&-|-\sqrt{\eta}\alpha\rangle_1|\sqrt{\eta}\alpha\rangle_2|-\sqrt{1-\eta}\alpha\rangle_{E1}
|\sqrt{1-\eta}\alpha\rangle_{E2}),
\end{eqnarray}
Tracing over all environment modes, we obtain the density operator:
\begin{align}
\varrho_{1,2}&=
\frac{1}{N_{-}}\{|\sqrt{\eta}\alpha\rangle_1\langle\sqrt{\eta}\alpha|\otimes|-\sqrt{\eta}\alpha\rangle_2\langle-\sqrt{\eta}\alpha|\nonumber\\
&+|-\sqrt{\eta}\alpha\rangle_1\langle-\sqrt{\eta}\alpha|\otimes|\sqrt{\eta}\alpha\rangle_2\langle\sqrt{\eta}\alpha|\nonumber\\
&- \Gamma_2(|\sqrt{\eta}\alpha\rangle_1\langle-\sqrt{\eta}\alpha|\otimes|-\sqrt{\eta}\alpha\rangle_2\langle\sqrt{\eta}\alpha|\nonumber\\
&+|-\sqrt{\eta}\alpha\rangle_1\langle\sqrt{\eta}\alpha|\otimes|\sqrt{\eta}\alpha\rangle_2\langle-\sqrt{\eta}\alpha|)\},
\end{align}
where $\Gamma_2=\exp\{-4(1-\eta)\alpha^2\}$.
Now it is interesting to note that in \cite{QIP2,decay} the authors have studied time evolution of ECSs within the master equation framework,
and the equivalent density operator was acquired (Corresponding to the noise parameter $\eta$, the effect of time
decay is characterized by $t\equiv\sqrt{\eta}=e^{-\gamma\tau/2}$, where $\tau$ is decoherence time and $\gamma$ is the decay
rate). Similar to case 1, first we give the "fully entangled fraction", which can also be comprehended as the
entanglement fidelity of the channel:
\begin{equation}
f(\varrho_{1,2})=\max_{\beta}\frac{(1+\Gamma_2)(e^{-2|\beta-\sqrt{\eta}\alpha|^2}+e^{-2|\beta+\sqrt{\eta}\alpha|^2}
-2e^{-|\beta-\sqrt{\eta}\alpha|^2-|\beta+\sqrt{\eta}\alpha|^2})}
{2(1-e^{-4|\alpha|^2})(1-e^{-4|\beta|^2})},
\end{equation}
To find the maximal value of $f(\varrho_{1,2})$, we investigate the derivative of $f(\varrho_{1,2})$ and find that
$\frac{d}{d\beta}f(\varrho_{1,2})=0$ when:
\begin{equation}
\beta=\sqrt{\eta}\alpha \,,
\end{equation}
So the entanglement fidelity of the channel is equal to:
\begin{equation}
f(\varrho_{1,2})=\frac{(1+e^{-4(1-\eta)|\alpha|^2})(1-e^{-4\eta|\alpha|^2})}
{2(1-e^{-4|\alpha|^2})} \,.
\end{equation}
In Fig. \ref{f2} we show the entanglement fidelity $f(\varrho_{1,2})$ as a function of $|\alpha|$ for
different parameter $\eta$. It is seen in Fig. \ref{f2} that in the limit $|\alpha|\rightarrow0$,
$f(\varrho_{1,2})\rightarrow\eta$.
However, as for the biphoton Bell state, the decohered density
operator is given as:
\begin{align}
\label{Bell2}
\varrho_{1,2}^{Bell}&=\eta^2|\psi^{Bell}\rangle_{1,2}\langle\psi^{Bell}|
+\frac{\eta(1-\eta)}{2}I_1\otimes|0\rangle_2\langle0|\hfill\nonumber\\
&+\frac{\eta(1-\eta)}{2}|0\rangle_1\langle0|\otimes I_2
+(1-\eta)^2|0\rangle_1\langle0|\otimes|0\rangle_2\langle0|,
\end{align}
Hence the entanglement fidelity is of order $\eta^2$ for the Bell state, which is clearly less than $\eta$ for
ECSs in the limit of small amplitudes. In the limit $|\alpha|\rightarrow\infty$, $f(\varrho_{1,2})\rightarrow1/2$
(except for $\eta=1$), and if $\eta=1/2$, the fidelity always equals to 1/2. Later we will show that in this limit
($|\alpha|\rightarrow\infty$) Alice and Bob actually share no entanglement.
\par
Before the discussion on decoherence of entanglement, we must emphasize that for a mixed state, there have been
many definitions for the measure of entanglement. In \cite{decay}, the authors have obtained the time evolution of entanglement
for an odd ECS using the definition described as \cite{PT}:
\begin{equation}
E=-2\sum_{i}\,\lambda_{i}^{-}
\end{equation}
where $\lambda_{i}^{-}$ are negative eigenvalues of $\varrho^{T_2}$ and $\varrho^{T_2}$ is
the partial transpose of the density operator. However, in this work, we still choose the definition
of entanglement of formation (EOF) and adopt Wootters's method \cite{EOF2,EOF3} to calculate the entanglement.
\par
\begin{figure}
\includegraphics[width=0.41\textwidth]{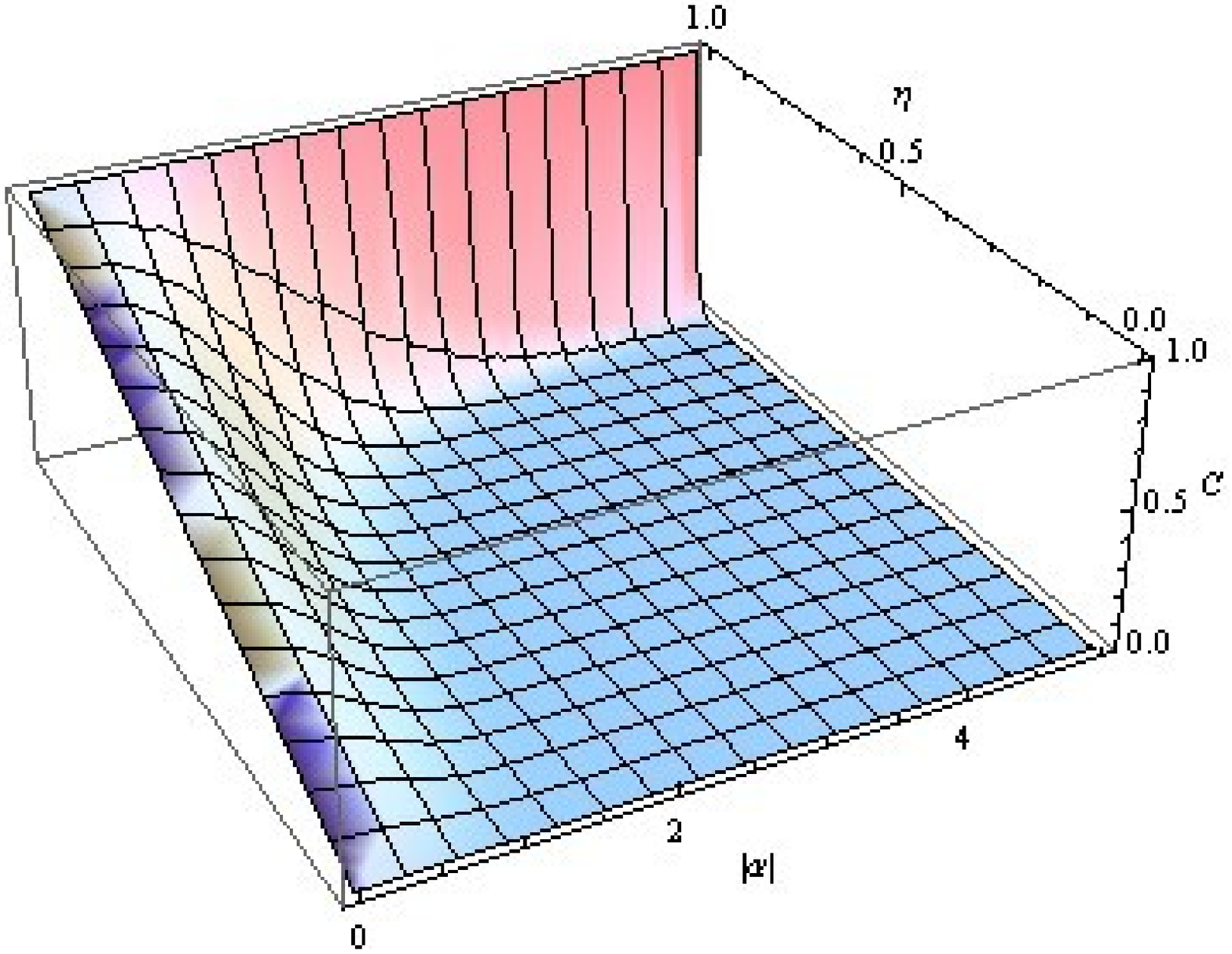}
$\quad$
\includegraphics[width=0.41\textwidth]{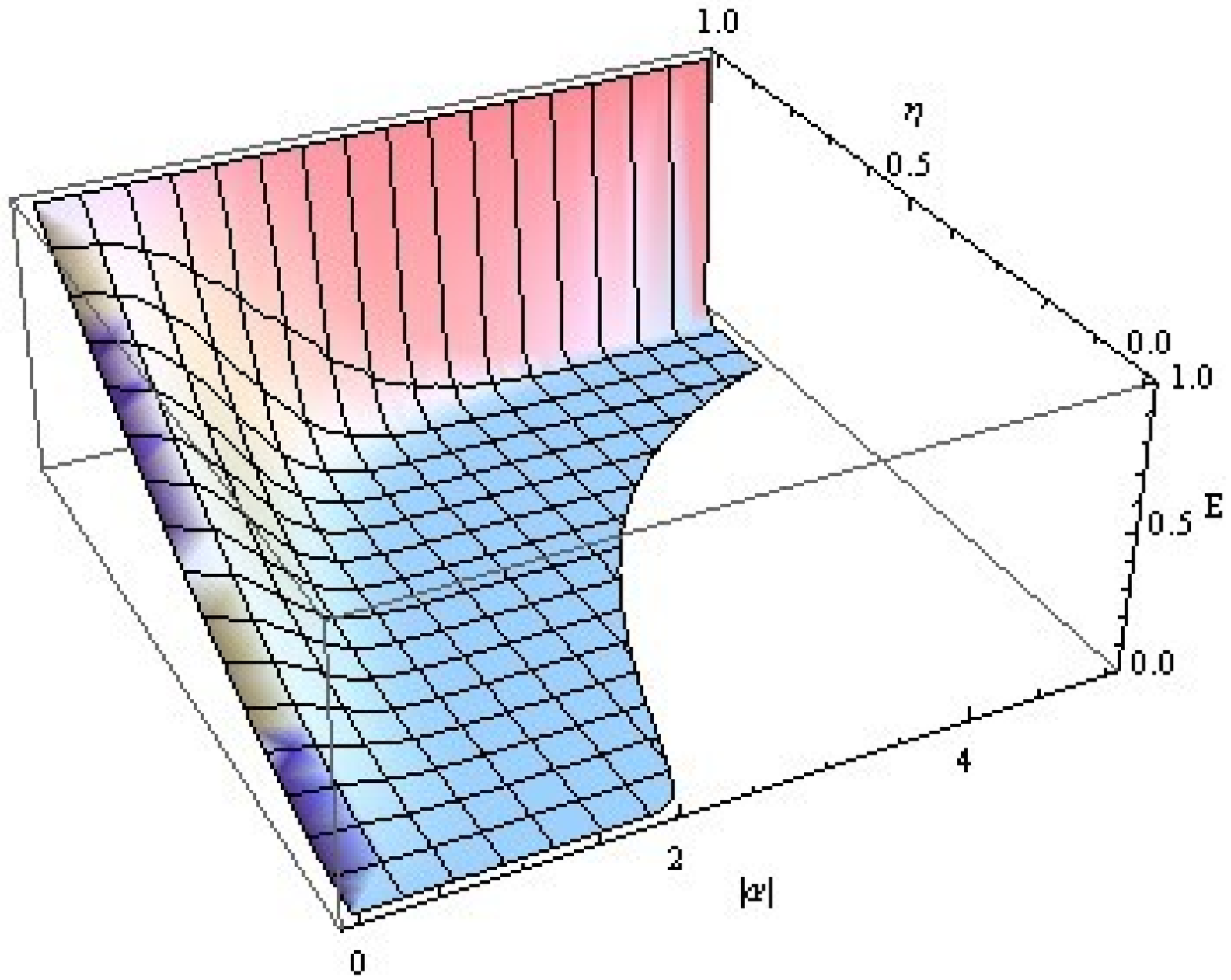}
\caption{The concurrence and EOF of $\varrho_{1,2}$ as a function of the coherent amplitude
$\alpha$ and the noise parameter $\eta$.}
\label{3D}
\end{figure}

For evaluation of entanglement, it is important to span the density operator on an orthogonal basis.
Though usually $|\alpha\rangle$ and $|-\alpha\rangle$ is nonorthogonal states, we can use the so called
even and odd coherent states defined as:
\begin{equation}
|\pm\rangle=\frac{1}{\sqrt{N_{\pm}(\eta)}}(|\sqrt{\eta}\alpha\rangle\pm|-\sqrt{\eta}\alpha\rangle),
\end{equation}
where $N_{\pm}(\eta)=2\pm2\exp(-2\eta\alpha^2)$. Although $|+\rangle$ and $|-\rangle$ are dependent
of $\eta$, they always remain orthogonal. The density matrix $\varrho_{1,2}$ is representing in the
orthonormal basis $\{|+,+\rangle,|+,-\rangle,|-,+\rangle,|-,-\rangle\}$ as:
\begin{equation}
\label{matrix}
\varrho_{1,2}=\frac{1}{16(1-e^{-4|\alpha|^2})}
\left(\begin{array}{cccc}
A & 0 & 0 & D \\
0 & B & -B & 0 \\
0 & -B & B & 0 \\
D & 0 & 0 & C
\end{array}\right),
\end{equation}
where A, B, C, D are defined as:
\begin{equation}
A=(1-\Gamma_2)N_{+}^{2}(\eta),
\end{equation}
\begin{equation}
B=(1+\Gamma_2)N_{+}(\eta)N_{-}(\eta),
\end{equation}
\begin{equation}
C=(1-\Gamma_2)N_{-}^{2}(\eta),
\end{equation}
\begin{equation}
D=-(1-\Gamma_2)N_{+}(\eta)N_{-}(\eta),
\end{equation}
\begin{equation}
\Gamma_2=\exp\{-4(1-\eta)\alpha^2\}.
\end{equation}
In section 1, we have showed that $\mathcal{E}(C)$ is a monotonous increasing function of the "concurrence"
C as C ranges from 0 to 1, so that one can take the quantity C as a kind of measure of entanglement in itself.
Following the general recipe \cite{EOF3}, the formula of concurrence is:
\begin{equation}
\label{C}
C=\max(0,\lambda_1-\lambda_2-\lambda_3-\lambda_4),
\end{equation}
where the parameters $\lambda_i\ (i=1,\ldots,4)$ are the square roots of the eigenvalues of the non-Hermitian
matrix $\varrho\tilde{\varrho}$, and
\begin{equation}
\tilde{\varrho}=(\sigma_y\otimes\sigma_y)\varrho^{\ast}(\sigma_y\otimes\sigma_y),
\end{equation}
Here $\varrho^{\ast}$ is the complex conjugate of $\varrho$ in the same basis (\ref{matrix}). Performing the
preceding calculations, one can get:
\begin{align}
\tilde{\varrho} = c_{0}\left(\begin{array}{cccc}
C & 0 & 0 & D \\
0 & B & -B & 0 \\
0 & -B & B & 0 \\
D & 0 & 0 & A
\end{array}\right), \,
\varrho\tilde{\varrho} = c_0^2\left(\begin{array}{cccc}
AC+D^2 & 0 & 0 & 2AD \\
0 & 2B^2 & -2B^2 & 0 \\
0 & -2B^2 & 2B^2 & 0 \\
2CD & 0 & 0 & AC+D^2
\end{array}\right)
\end{align}
where $c_0=\frac{1}{16}\{1-\exp(-4\alpha^2)\}$. The eigenvalues of $\varrho\tilde{\varrho}$ are
$c_0^2\{4B^2,AC-2\sqrt{AC}D+D^2,AC+2\sqrt{AC}D+D^2,0\}$. Note that $\sqrt{AC}=-D$, so the square roots are
$c_0\{2B,-2D,0,0\}$. The concurrence of $\varrho$ is given by:
\begin{equation}
\label{Concurrence}
C=c_0\max(0,2B+2D)=\frac{e^{4\eta\alpha^2}-1}{e^{4\alpha^2}-1}\in[0,1] \ (for\, \eta\in[0,1]) \,.
\end{equation}
Therefore the entanglement of formation of $\varrho$ is just
$E(\varrho)=\mathcal{E}(C)=H(\frac{1}{2}+\frac{1}{2}\sqrt{1-(\frac{e^{4\eta\alpha^2}-1}{e^{4\alpha^2}-1})^2})$.
The concurrence and EOF of $\varrho$ is shown in Fig. \ref{3D}.
From this figure one can see that for $|\alpha|\rightarrow0$, $C\rightarrow\eta$. For $\eta=1$,
C is always equal to 1. In the limit $|\alpha|\rightarrow\infty$, the concurrence approaches
0 (for $\eta<1$). Now we will demonstrate that in this limit ($|\alpha|\rightarrow\infty$)
Alice and Bob actually share no entanglement.
\par
Inspired by the works in \cite{decay}, we consider the Peres Criterion: the necessary and sufficient condition
for separability of a two-dimensional bipartite system is the positivity of the partial transposition
of its density matrix \cite{separability}. The partial transpose of $\varrho_{1,2}$ is:
\begin{equation}
\varrho^{T_2}=c_0
\left(\begin{array}{cccc}
A & 0 & 0 & -B \\
0 & B & D & 0 \\
0 & D & B & 0 \\
-B & 0 & 0 & C
\end{array}\right),
\end{equation}
The four eigenvalues of $\varrho^{T_2}$ are as follows: $c_0\{\frac12\left(A+C\pm\sqrt{A^2+4B^2-2AC+C^2}\right),
B\pm D\}$. For $|\alpha|\rightarrow\infty$, $\Gamma_2\rightarrow0$, and thus $A=B=C=-D>0$. One can easily verify that
all the four eigenvalues are nonnegative. Hence Alice and Bob actually obtain a separable state in this limit.
\par
\begin{figure}
\includegraphics[width=0.41\textwidth]{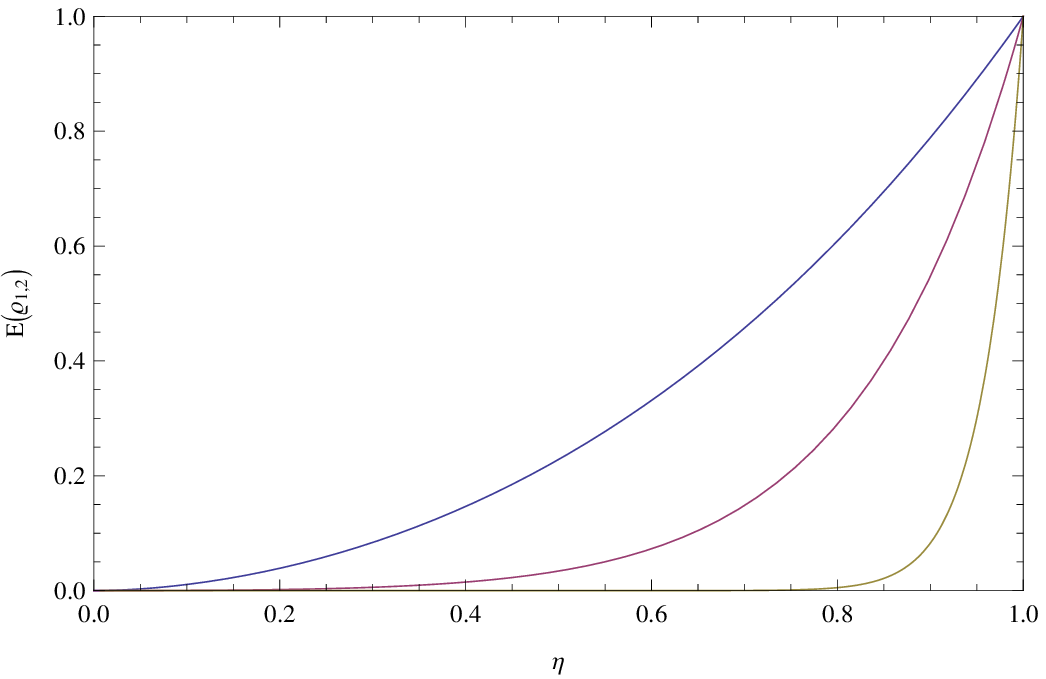}
$\quad$
\includegraphics[width=0.41\textwidth]{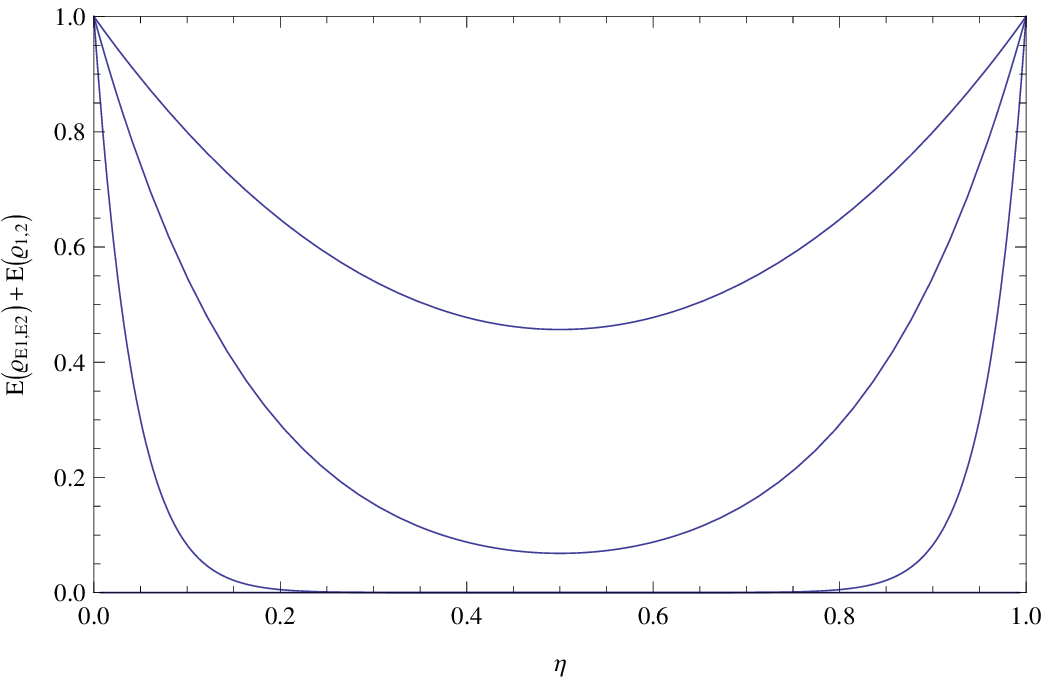}
\caption{The EOF of $\varrho_{1,2}$ and $E(\varrho_{1,2})+E(\varrho_{E1,E2})$ ($E(\rho_{E1,E2})$ denotes
the entanglement between environment mode E1 and E2, which can be easily obtained in a similar way to $E(\varrho_{1,2}$))
as a function of the noise parameter $\eta$ for several coherent amplitudes $\alpha=0.5,1.0,2.0$ from top to bottom.}
\label{E2}
\end{figure}
\begin{figure}
\begin{center}
\includegraphics[width=0.41\textwidth]{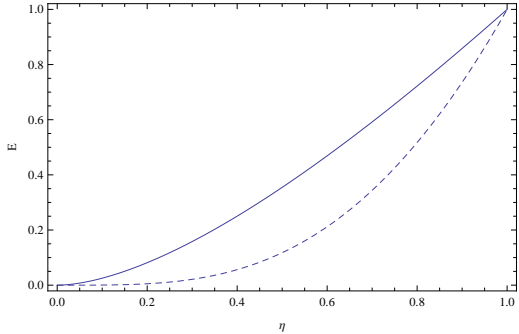}
\end{center}
\caption{The entanglement of formation of $\varrho_{1,2}$ (solid line) and $\varrho_{1,2}^{Bell}$ (dashed line)
as a function of $\eta$ in the limit $\alpha\rightarrow0$.}
\label{limit}
\end{figure}

Finally, for comparison, we plot the degrees of entanglement for ECSs for several values of $\alpha$ in
Fig. \ref{E2} in this symmetric noise channel case. Now we are fully convinced that the ECSs with large
coherent amplitudes decohere faster than those with small amplitudes.
Correspondingly, note that at the right side of Eq. (\ref{Bell2}) there are four terms: the first one is just
a Bell state (two-photon state), while the other three terms are simply separable states (single-photon state or
completely vacuum state).
Therefore the concurrence is:
\begin{equation}
C(\varrho^{Bell})=\eta^2 \,.
\end{equation}
This result is equivalent to the Eq. (9) in \cite{decay}. However, in the limit of $|\alpha|\rightarrow0$, we already
know $C(\varrho)\rightarrow\eta>\eta^2$ for $0<\eta<1$. Since $\mathcal{E}(C)$ is a monotonous increasing function
of the concurrence, it is worth noticing that $E(\varrho)>E(\varrho^{Bell})$ in this limit. Note that this result
is just opposite to the Eq. (10) ($E_{ECS}<E_{EPP}$, EPPs stand for entangled photon pairs) in \cite{decay}, thus the conclusion that
"Obviously, the EPP is always more entangled than the ECS for any values of $\alpha$" \cite{decay} can \textbf{not} be drawn
from our analysis by using the definition of EOF. We have plotted the entanglement decoherence for the ECSs and
bipartite Bell states in the limit of $\alpha\rightarrow0$ in Fig. \ref{limit}.
\section{Conclusion Remarks}
We have investigated the effects of channel decoherence on the ECSs by utilizing the concept
of "entanglement of formation". In our work, two situations are considered: (1) asymmetric
noise channel and (2) symmetric noise channel. In case (1), we have given the "fully entangled fraction" $f(\rho)$
(which can also be comprehended as the entanglement fidelity of the channel) of the decohered ECS $\rho$, and
based on that we have calculated the lower bound of EOF of $\rho$. In case (2), we have made a further discussion
on decoherence properties of the mixed ECS $\varrho$. Not only have we obtained the entanglement fidelity
of the mixed ECS $\varrho$, but we have found the explicit expressions of the "concurrence" $C(\varrho)$ and the
EOF $E(\varrho)$. There are two points which we need take notice in this case:\\
\textbf{Ramark 1:} In the limit of $|\alpha|\rightarrow\infty$, we see from Fig. \ref{f2} that the fidelity $f(\varrho)$ approaches
1/2 for any $\eta<1$. According to \cite{Horodecki}, in the standard teleportation scheme, the maximal fidelity achievable
from a given bipartite state $\varrho$ is:
\begin{equation}
F=\frac{2f+1}{3} \,,
\end{equation}
where $f$ is the fully entangled fraction. It means that for $|\alpha|\gg1$, our result gives $F=2/3$,
which is just the classical limit of teleportation fidelity. In fact, we have shown that Alice and Bob share no entanglement
when $|\alpha|\rightarrow\infty$.
\\
\textbf{Ramark 2:} In the limit of $|\alpha|\rightarrow0$, the authors of \cite{decay}
found that $E_{ECS}<E_{EPP}$ following their preceding entanglement
definition \cite{PT} and thus concluded the EPP (entangled photon pair) is more entangled than the ECS for any values of $\alpha$.
However, by employing the definition of EOF, we have shown $C(\varrho)=\eta>C(\varrho^{Bell})=\eta^2$ in this limit.
Since $\mathcal{E}(C)$ is monotonically related to C, one can easily see $E_{ECS}>E_{Bell}$ as shown in Fig.
\ref{limit}.
\par
In fact, the definition in \cite{PT} is just a kind of measure of entanglement called "\textbf{Negativity}"
\cite{negativity1,negativity2,negativity3}. In Ref. \cite{negativity2} the authors studied a interesting theme:
the ordering of the density operator with respect to the degree of entanglement using different measurements.
Specifically, one would expect that any two "reasonable" entanglement measures should give the same ordering of
the density operators, which means for certain two entanglement measures $E_1$ and $E_2$, and arbitrary two density operators
$\rho_1$ and $\rho_2$, we will have the relationship:
\begin{equation}
\label{condition}
E_1(\rho_1)>E_1(\rho_2)\,\Leftrightarrow\,E_2(\rho_1)>E_2(\rho_2),
\end{equation}
However, for the decohered ECS and Bell states, in the limit of $\alpha\rightarrow0$
our result combined with that of \cite{decay} reveals that:
\begin{equation}
E_F(\varrho^{ECS})>E_F(\varrho^{Bell})\,\Leftrightarrow\,E_N(\varrho^{ECS})<E_N(\varrho^{Bell}),
\end{equation}
where $E_F$ and $E_N$ denote "entanglement of formation" and "negativity", respectively. This happens to be
a counterexample which violates the condition (\ref{condition}). For further details, one can see \cite{ordering1,ordering2}
for a review.
\par
To sum up, we have found that the larger initial coherent amplitude $\alpha$, the faster ECSs decohere. Besides,
our work shows that ECSs with sufficient small $\alpha$ are more robust against channel decoherence than biphoton entangled states
in both cases.
\section*{Acknowledgements}
This work was supported by the National Basic Research Program of China (Grants No. 2011CBA00200 and No. 2011CB921200),
National Natural Science Foundation of China (Grant NO. 60921091), and China Postdoctoral Science Foundation (Grant No. 20100480695).

\end{document}